# Construction of various time-dependent Hamiltonians on a single photonic chip

Rui Ye[1,†], Guangzhen Li[1,†,*], Shuai Wan[2,3,†], Xiaotian Xue[1,†], Piyu Wang[2,3,†], Xin Qiao[4], Hao Li[1], Shijie Liu[1], Jiayu Wang[1], Rui Ma[5], Fang Bo[5], Yuanlin Zheng[1,6,*], Chunhua Dong[2,3,*], Luqi Yuan[1,*], and Xianfeng Chen[1,6,7,*]

[1]State Key Laboratory of Advanced Optical Communication Systems and Networks, School of Physics and Astronomy, Shanghai Jiao Tong University, Shanghai 200240, China

[2]CAS Key Laboratory of Quantum Information, University of Science and Technology of China, Hefei, Anhui 230026, China

[3]CAS Center for Excellence in Quantum Information and Quantum Physics, University of Science and Technology of China, Hefei, Anhui 230088, China

[4]College of Physics and Electronics Engineering, Northwest Normal University, Lanzhou 730070, China

[5]MOE Key Laboratory of Weak-Light Nonlinear Photonics, TEDA Applied Physics Institute and School of Physics, Nankai University, Tianjin 300457, China

[6]Shanghai Research Center for Quantum Sciences, Shanghai 201315 China

[7]Collaborative Innovation Center of Light Manipulations and Applications, Shandong Normal University, Jinan 250358, China

*Corresponding authors: liguangzhen520@sjtu.edu.cn; ylzheng@sjtu.edu.cn; chunhua@ustc.edu.cn; yuanluqi@sjtu.edu.cn; xfchen@sjtu.edu.cn

†These authors contributed equally to this work.

**Abstract**

Integrated photonics provides an important platform for simulating physical models with high-performance chip-scale devices, where the lattice size and the time-dependence of a model are key ingredients for further enriching the functionality of a photonic chip. Here, we propose and demonstrate the construction of various time-dependent Hamiltonian models using a single microresonator on thin-film lithium niobate chip. Such an integrated microresonator holds high quality factor to $10^6$, and supports the construction of the synthetic frequency lattice with effective lattice sites up to 152 under the electro-optic modulation. By further applying a bichromatic modulation composed of two radio-frequency signals oppositely detuned from the resonant frequency in the microresonator, we build different time-dependent Hamiltonians with the time-varying nearest-neighbor coupling strength in synthetic frequency lattice. We measure the temporal features from capturing the dynamic band structures of the lattice and demonstrate a variety of time-dependent synthetic lattice models by engineering the driven pattern of the modulation, highlighting great flexibility of the microresonator. Our work shows a photonic chip for simulating versatile time-dependent Hamiltonians, which pushes forward quantum simulations in integrated photonics with great experimental tunability and reconfigurability.

**Introduction**

Recent developments in photonic chips have exhibited their great capability for achieving versatile physical models with highly compact integration and small footprint size[1–3], and enabling novel nanophotonic functionalities with programmable photonic circuits[4–6] towards quantum computing[7–10], high-efficiency frequency conversion[11,12] and high-bandwidth modulation[13,14]. For these on-chip photonic devices, it usually desires the physical model with the large scale and the on-demand tunability to trigger more opportunities for the integrated photonic applications, where the adjustment can be achieved by different fabrications for specific functionality[15], controlled by patterned pump beam[16], or through cascading electro-optically and thermo-optically controlled elements[4,17–19]. Although programmable waveguide arrays are able to precise control independent Hamiltonian terms[20], these present configurations hold time-independent Hamiltonian features and are limited to static physical models. Recent updates on curved waveguides on photonic chips mimic the time-dependent evolution of the model along the waveguide direction[21], but the lattice site number is limited due to the small footprint and the device is lack of temporal tunability. On the other hand, periodically driven physical systems characterized by time-dependent Hamiltonians (i.e., Floquet system), hold profoundly distinctive features and thus inspire strong study interests[22–25], with exotic photonic phenomena including photon-assisted tunneling[26], Floquet topological insulators[27,28], Floquet PT-symmetry[29] and Floquet-Bloch osillations[30]. Therefore, it is of fundamental importance in realizing the time-dependent Hamiltonian in the integrated photonics, supporting the physical model with a large site number and on-demand temporal tunability, which may offer more possibilities for investigating dynamics of complex physical systems on optical platforms.

The recent surge of interest in synthetic frequency dimensions brings more opportunities to integrated photonics for exploring higher-dimensional physics in simple structures with lower-dimensional geometry[31–39]. The constructed physical models in the synthetic frequency dimension depend on flexible external modulation, enabling the realizations of rich exotic phenomena across many photonic areas, including programming photonic simulator[36], quantum walk comb[37], coherently driven dissipative solitons[38], and extraction of the topological invariant[39]. Taking advantage of the high quality ($Q$) factor, low loss and dense integration of chip-scale devices based on the lithium niobite on insulator technology[1,40,41], synthetic frequency dimension has also been implemented on photonic chips[42–45], demonstrating interesting physics including random walks and Bloch oscillations[42], mirror-induced reflection[43], quantum simulators[44], and arbitrarily reconfigurable frequency lattices[45]. The good side for achieving synthetic frequency dimension in integrated photonics lies on the large site number in the physical model and the extra tunability provided by strong and fast responses of the platform material to the external modulation. However, up to date, the thorough demonstration for constructing various time-dependent Hamiltonians with synthetic frequency dimension on a single photonic chip has not been fully performed.

In this work, we show the experimental realization of constructing various time-dependent Hamiltonians in a high-$Q$ microresonator fabricated on a thin-film lithium niobate (TFLN) chip. By applying a bichromatic near-resonant electro-optic (EO) modulation with two modulation frequencies oppositely detuned from the resonant frequency, a periodically driven synthetic lattice model with time-varying coupling strength is constructed in the synthetic frequency dimension. Such synthetic lattice supports a large number of sites up to 152. We capture the temporal features of the synthetic lattice by measuring the trajectories of the dynamic band structures, where the time-varying coupling strength is manipulated by tuning the external driving signal. We also showcase the capability of the fabricated tunable device for constructing versatile time-dependent Hamiltonians by engineering the modulation patterns. Our study presents a proof-of-principle demonstration on simulating the time-dependent Hamiltonians in a single on-chip microresonator associated with synthetic frequency dimension, which holds great potentials in the future simulations of periodically driven non-Hermitian models with the possible extension to higher dimensions in the integrated photonics[46–49].

**Results**

We demonstrate the configurable construction of synthetic frequency lattice supporting various time-dependent Hamiltonian models in a racetrack microresonator coupled with a bus waveguide as sketched in Fig. 1a, fabricated on the x-cut TFLN photonic chip [see Methods]. By ignoring the managed weak group velocity dispersion, the microresonator supports a set of equally-spaced resonant modes with frequencies $\omega_n = \omega_0 + n\Omega_R$, where $\Omega_R$ is the free spectral range (FSR) of the microresonator, $\omega_0$ is a reference frequency, and $n$ is the index for the $n$th mode. A pair of electrodes are placed on both sides of the microresonator, so EO phase modulation with modulation frequency $\Omega_M$ can be applied on the microresonator in the microwave regime. By applying resonant modulation $g_0\cos(\Omega_M t)$ with $\Omega_M \approx \Omega_R$, one can couple the adjacent frequency modes together, and then form a one-dimensional synthetic lattice along the frequency axis of light. Under the resonant case ($\Omega_M = \Omega_R$), the model gives the energy band as $\varepsilon_{k_f} = g_0\cos(k_f\Omega_R)$, which holds a static band feature with constant nearest-neighbor (NN) coupling strength $g_0$. Here, $k_f$ denotes the wave vector reciprocal to the frequency dimension, which in unit is a time variable[33,35]. Different with previous works[33–39], we verify the capability for studying time-dependent Hamiltonian in this on-chip device. Therefore, we consider a bichromatic near-resonant modulation in the form of

$$J(t) = g_1\cos[(\Omega_R - \Delta_1)t] + g_2\cos[(\Omega_R + \Delta_2)t + \phi], \quad (1)$$

which composes of two modulation signals with modulation frequencies $\Omega_{M1} = \Omega_R - \Delta_1$ and $\Omega_{M2} = \Omega_R + \Delta_2$, oppositely detuning from the resonant frequency $\Omega_R$. Constant parameters $g_1, g_2$ and $\phi$, are the corresponding modulation strengths and phase difference between two signals, respectively. Such a bichromatic modulation can be achieved by applying the composition of two radio frequency (RF) signals on the patterned electrodes through a microwave coupler. Then one obtains the Hamiltonian of the system in Fig. 1a as

$$H = \sum_n \omega_n a_n^\dagger a_n + J(t)\sum_n (a_{n+1}^\dagger a_n + a_n^\dagger a_{n+1}), \quad (2)$$

with $a_n^\dagger$ ($a_n$) being the creation (annihilation) operator for the $n$th mode. As the first proof-of-principle demonstration, we set the modulation signal in Eq. (1) to be $g_1 = g_2 = g$, $\Delta_1 = \Delta_2 = \Delta$. After transferring the Hamiltonian into the interaction picture and taking the rotating-wave approximation, we obtain a simplified Hamiltonian

$$H_C = \sum_n g\cos(\Delta \cdot t + \phi/2)[c_{n+1}^\dagger c_n e^{-i\phi/2} + c_n^\dagger c_{n+1} e^{i\phi/2}], \quad (3)$$

with $a_n = c_n e^{-i\omega_n t}$. One can see that the Hamiltonian in Eq. (3) describes a tight-binding lattice model but with NN coupling being time-dependent, where we can use $G(t) = g\cos(\Delta \cdot t + \phi/2)$ to label such time-varying coupling strength in the synthetic frequency lattice here.

Such a time-dependent Hamiltonian holds fundamental difference from static Hamiltonian in quantum mechanisms. To see such difference, we convert the Hamiltonian in Eq. (3) into the $k_f$ space at each time slice and obtain the instantaneous band structure of the system as

$$\varepsilon_{k_f}(t) = 2g\cos\left(\Delta \cdot t + \frac{\phi}{2}\right) \cdot \cos\left(k_f \Omega_R + \frac{\phi}{2}\right) = 2G(t) \cdot \cos\left(k_f \Omega_R + \frac{\phi}{2}\right). \quad (4)$$

One can see that for a time-dependent model, the eigenvalue is not constant so the band is varying with the time evolution. In particular, such band exhibits the cosinusoidal-shape band feature while the amplitude of the band periodically oscillates over time due to the time-varying coupling strength $G(t)$ [see Fig. 1b].

To explore such a time-dependent Hamiltonian, Floquet theory is usually taken to do the analysis[22–25]. Once the Floquet analysis is applied on the constructed time-dependent lattice in Fig. 1a, it results in time-independent bands, which are completely flat [see Fig. S1]. As an alternative way to prove the construction of a time-dependent model, the dynamic evolution of the wave vector reciprocal to the frequency axis of the excitation in the synthetic frequency lattice can be taken, where instead of studying the evolution of the wave, one tracks the evolution of the band structure in time using the dynamic band structure method proposed in Ref. [35]. The main idea is that the system is excited by a pump light with the linearly varying frequency $\omega(t)$ which hits the dynamic band structure in each time slice, i.e., $\varepsilon_{k_f}(t_m) = \omega(t_m) = 2\pi\eta t_m$, where $\eta$ denotes the change rate on the frequency of the source and $m$ is an integer. At any given time slice $t_m$, the corresponding frequency $\omega(t_m)$ can excite the corresponding eigenvalues on the band $\varepsilon_{k_f}(t_m)$, which are two points at wave vectors $k_{f,m1}$ and $k_{f,m2}$ satisfying Eq. (4). By connecting $[k_{f,m1(2)}, \varepsilon_{k_f}(t_m)]$ for all time slices, one obtains the trajectory of the dynamic band structure as illustrated in Figs. 1b-1c.

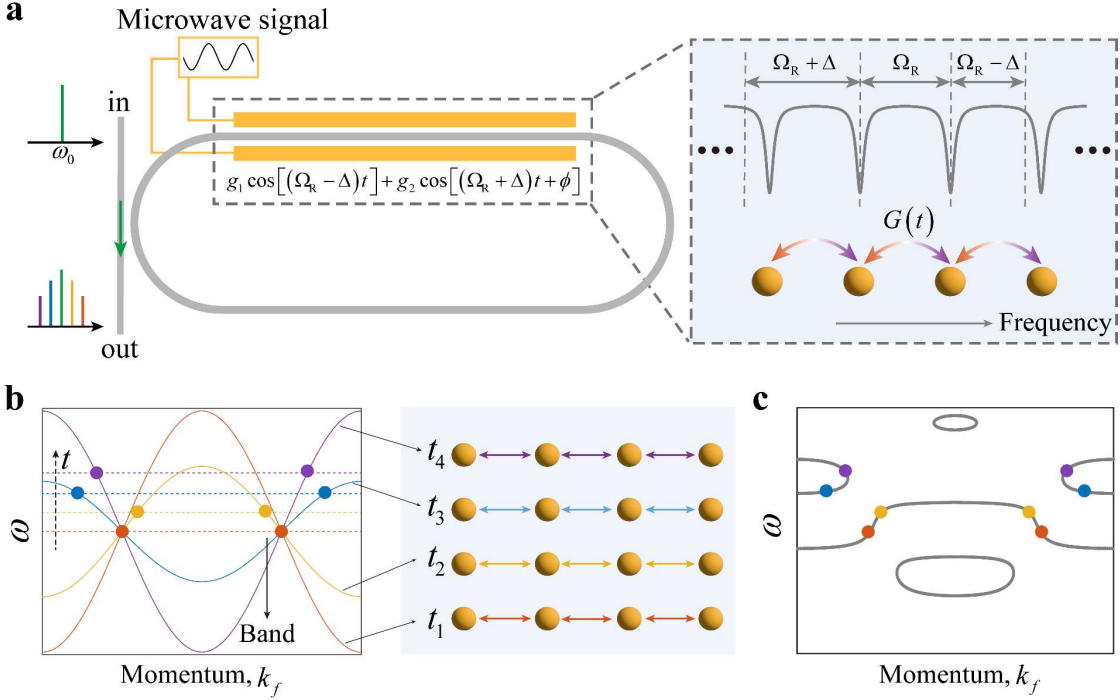

**Fig. 1 Configuration of periodically driven photonic lattice in synthetic frequency dimension. a**, A racetrack microresonator with patterned electrodes is driven by a bichromatic near-resonant modulation, composed of two RF signals with their frequencies oppositely detuned from the resonant frequency $\Omega_R$. The yellow spheres denote the constructed periodically driven lattice model along the frequency dimension, with time-varying coupling strength $G(t)$ labeled by the color-gradient arrows. **b**, Schematic evolution of the static band structures of the system in **a** with time varying from $t_1$ to $t_4$. The intersection points of the excitation frequency $\omega(t_m)$ (dashed lines) and the corresponding band structure $\varepsilon_{k_f}(t_m)$ at different times $t_m$ are labeled by the colored circles. The right column represents the synthetic lattice models under time-varying coupling strength determined by $G(t)$. **c**, Schematic trajectory of the dynamic band structure, formed by the intersection points in **b** for all the time slices.

In experiments, we fabricate the racetrack microresonator on TFLN chip with a circumference of 26.5 mm [see Fig. 2a and Fig. S2]. To characterize the steady-state response of the fabricated device, we first measure the transmittance spectrum of the microresonator by linearly sweeping the wavelength of the pump light spanning the telecom bands [see Methods for experimental details]. We obtain a loaded $Q$ factor of $Q_L = 1.23 \times 10^6$ and measured

FSR of $\Omega_R = 4.955$ GHz (see Fig. 2b and Fig. S2a-S2b). Broadband EO frequency comb generation can be achieved by dispersion engineering[50,51]. The system exhibits nearly flat dispersion profile for the fundamental TE mode at around 1568 nm [see Methods], and the measured second-order group velocity dispersion parameter of the microresonator is $D_2/2\pi = -6$ kHz [see Fig. 2c]. We then excite the microresonator at the wavelength of 1568 nm (corresponding to $\omega_0 = 191.194$ THz) the zero-dispersion point. An optical frequency comb with a frequency spanning of 753 GHz, up to 152 resonant modes, is obtained as plotted in Fig. 2d. This broadband frequency spanning proves the effective hopping between the frequency lattice sites and therefore the successful construction of the chip device for simulating models in the synthetic frequency dimension in experiment.

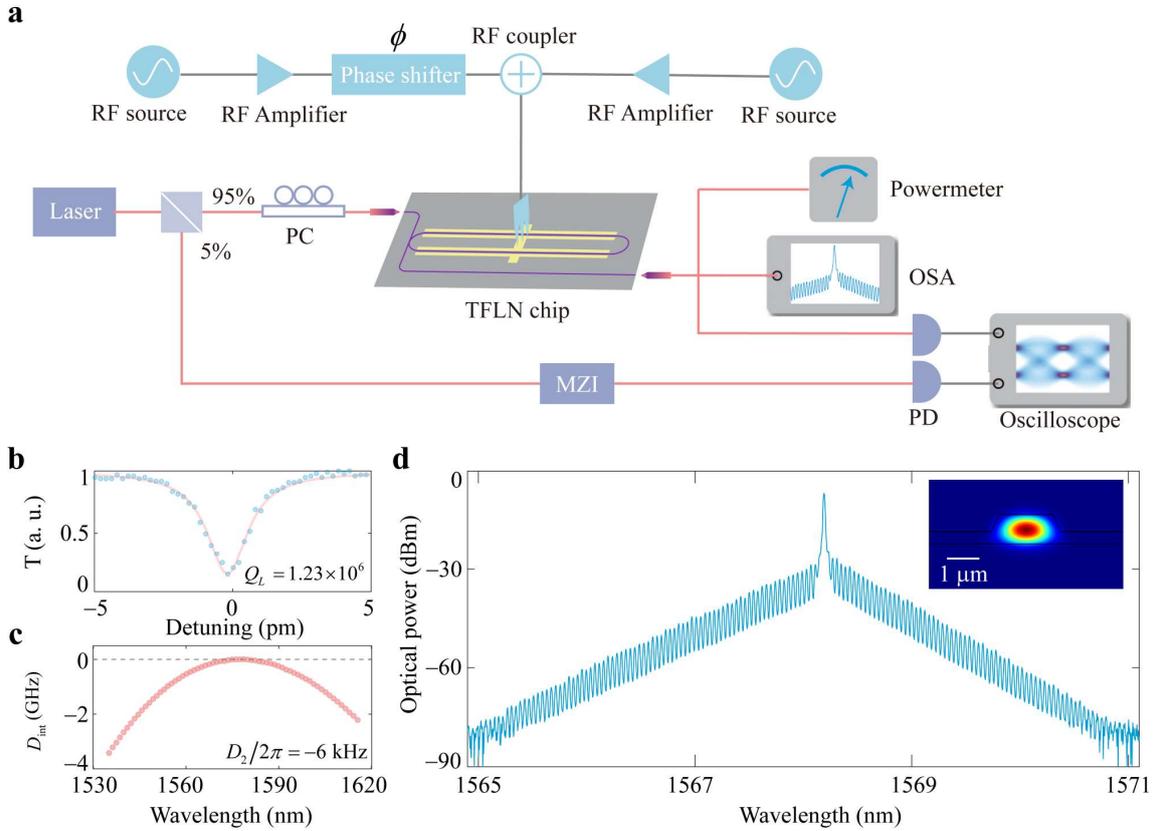

**Fig. 2. a**, Experimental setup. PC, polarization controller; OSA, optical spectrum analyzer; MZI, Mach–Zehnder interferometer; PD, photodiode. **b**, Transmission spectrum at the resonant length of 1568 nm revealing a high loaded $Q$ factor of the microresonator. **c**, Measured integrated dispersion ($D_{int}$) as a function of the pump wavelength. **d**, The spectrum of the EO frequency comb generated in the microresonator. The inset shows the simulated profiles (amplitude of the horizontal electric-field component) of the fundamental TE mode in the nanowaveguide.

To construct the time-dependent Hamiltonian in the synthetic frequency lattice illustrated in in Fig. 1 on the fabricated device, one needs to engineer the time-varying coupling strength $G(t)$ in experiments. A bichromatic EO modulation signal with an experimental form of $V(t) = V_0\cos(\Omega_{M1}t) + V_0\cos(\Omega_{M2}t + \phi)$ is then applied on the patterned electrodes of the chip [see Fig. 2a]. The phase difference $\phi$ between the two RF generators is phase-locked using an internal 10-MHz clock. To unveil the dynamics in the built synthetic lattice, we perform the dynamic band structure measurement by using the time-resolved transmission spectroscopy[33,35], which is also suitable for the near-resonant case. To do so, we finely scan the frequency of the pump laser centered at 1568 nm across several FSRs of the microresonator, and record the corresponding transmission spectrum. The collected transmission spectrum is sliced with fixed time window, in which way one can obtained the time-resolved band structures of the system after vertically stacking these time windows as a function of the input frequency $\omega(t)$.

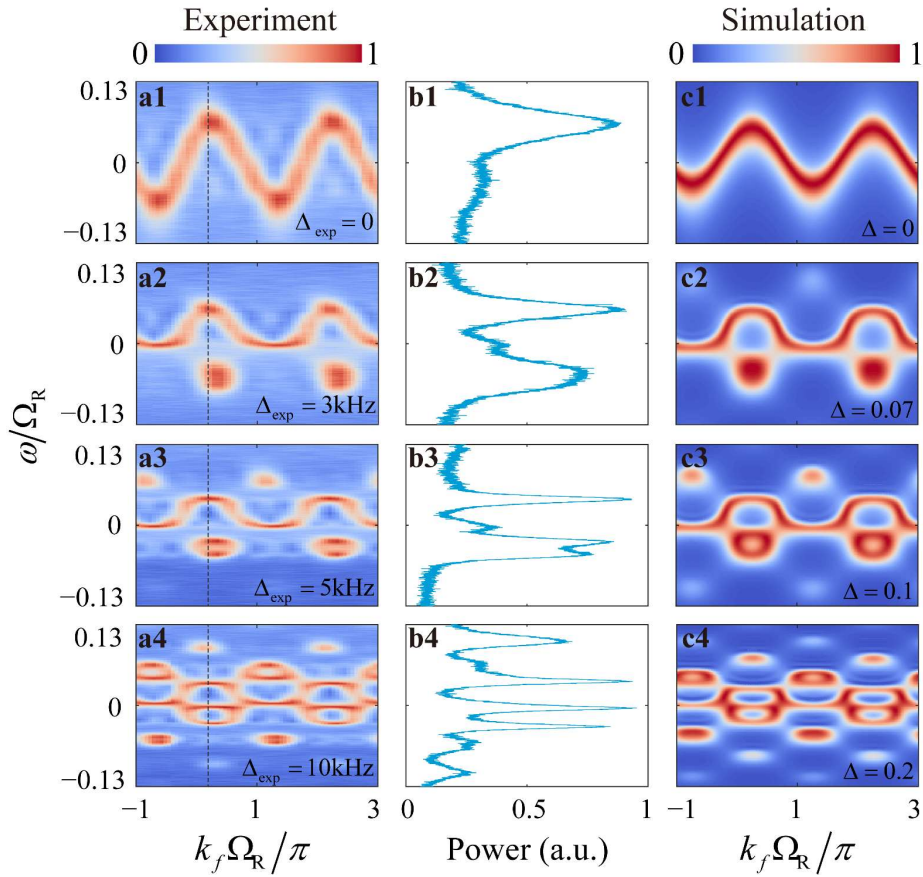

**Fig. 3 a1-a4**, Measured trajectories of the dynamic band structures under modulation $V(t) = V_0\cos[(\Omega_R - \Delta_{\exp})t] + V_0\cos[(\Omega_R + \Delta_{\exp})t + \phi]$, with varied experimental frequency detunings $\Delta_{\exp}$ and $V_0 = 2.5$ V, $\phi = 1.5\pi$. **b1-b4**, Vertically normalized power distributions at the position $k_f\Omega_R = 0.2\pi$, labeled by the dashed lines in **a1-a4**. **c1-c4**, Simulated dynamic band structures under modulation $J(t) = g\cos[(\Omega_R - \Delta)t] + g\cos[(\Omega_R + \Delta)t + \phi]$, with different theoretical frequency detunings $\Delta$, and $g = 0.4$, $\phi = 1.5\pi$, $\eta = 0.002$.

By taking modulation frequencies $\Omega_{M1} = \Omega_R - \Delta_{\exp}$ and $\Omega_{M2} = \Omega_R + \Delta_{\exp}$, we build the synthetic frequency lattice in Eq. (3) with different time-dependent $G_{\exp}(t) = V_0\cos(\Delta_{\exp}t + $ $\phi/2$) and vary $\Delta_{\text{exp}}$ from 0, 3 kHz, 5 kHz to 10 kHz. Such choices of parameters regard to the same modulation frequency detunings ($\Delta_{\text{exp}}$) in the two RF signals but opposite detuned from the resonant frequency $\Omega_R$, which gives a static lattice with $\Delta_{\text{exp}} = 0$ but different time-varying periodicity for the other three cases. The measured trajectories of the corresponding band structure in different synthetic lattices are shown in Figs. 3a1-3a4. Note that we slice the transmission spectrum by two roundtrip times of the microresonator, which gives the horizontal periodicity of the band structure $k_f \Omega_R \in [-\pi, 3\pi]$. When the system is static, i.e., $\Delta_{\text{exp}} = 0$ (corresponding to the resonant modulation case where $\Omega_{M1} = \Omega_{M2} = \Omega_R$), the band structure exhibits a cosinusoidal shape [see Fig. 3a1]. The dynamic feature manifests once the time-varying $G_{\text{exp}}(t)$ is built ($\Delta_{\text{exp}} \neq 0$), where the measurements give splitting patterns as shown in Figs. 3a2-a4. The trajectory of the dynamic band structure associated to different time-dependent Hamiltonians is along the vertical energy axis, due to the oscillation rate of $G_{\text{exp}}(t)$ inversely proportional to the frequency detuning $\Delta_{\text{exp}}$. It also reflects the stationary wave feature of the band structure due to the time-varying coupling strength. The splitting band number increases with the frequency detuning while the vertical energy window remains unchanged. Such phenomena correspond to the physics picture that the photon oscillates faster at fixed $k_f$ position in the synthetic lattice due to the coupling strength $G(t)$ with the larger oscillation rate, as seen in the power distributions for the field sliced at $k_f \Omega_R = 0.2\pi$ in Figs. 3b1-b4. We also simulate the dynamic band structures by solving the Hamiltonian in Eq. (3) numerically [see Figs. 3c1-c4], which gives the resulting trajectories agreeing well with the experimental results.

In a conventional time-dependent Hamiltonian, the phase information in the coupling strength $G(t)$ does not affect the physical observable. Nevertheless, the synthetic frequency lattice here provides a way to extract the phase information in the coupling strength due to the synchronization with the phase influencing the static band part $\cos(k_f \Omega_R + \phi/2)$ in Eq. (4) simultaneously. We explain this simple physical picture using the theoretical static band structures at different time slices $(t_1, t_2, t_3, t_4)$ with the fixed $\Delta = 0.1$ and the varying phase difference $\phi$ in Figs. 4a1-a4. We see that the bands evolve along the horizontal $k_f$ direction when varying $\phi$ with a periodicity $2\pi$ and the shape of $G(t)$ is also modified by $\phi$ at different time slices. In experiments, we fix $\Delta_{\text{exp}} = 5$ kHz and change $\phi$ from $0, 0.5\pi, 1\pi$, to $1.5\pi$. The measured dynamic band structures are plotted in Figs. 4b1-b4, which verifies the theoretical prediction and also matches well with numerical simulations in Figs. 4c1-c4. The measured bands in Figs. 3 and 4 indicate that our experimental setup exhibits relatively low loss benefiting from the high $Q$ factor of the microresonator, which provides the capability for the on-demand photonic simulations of time-dependent Hamiltonians.

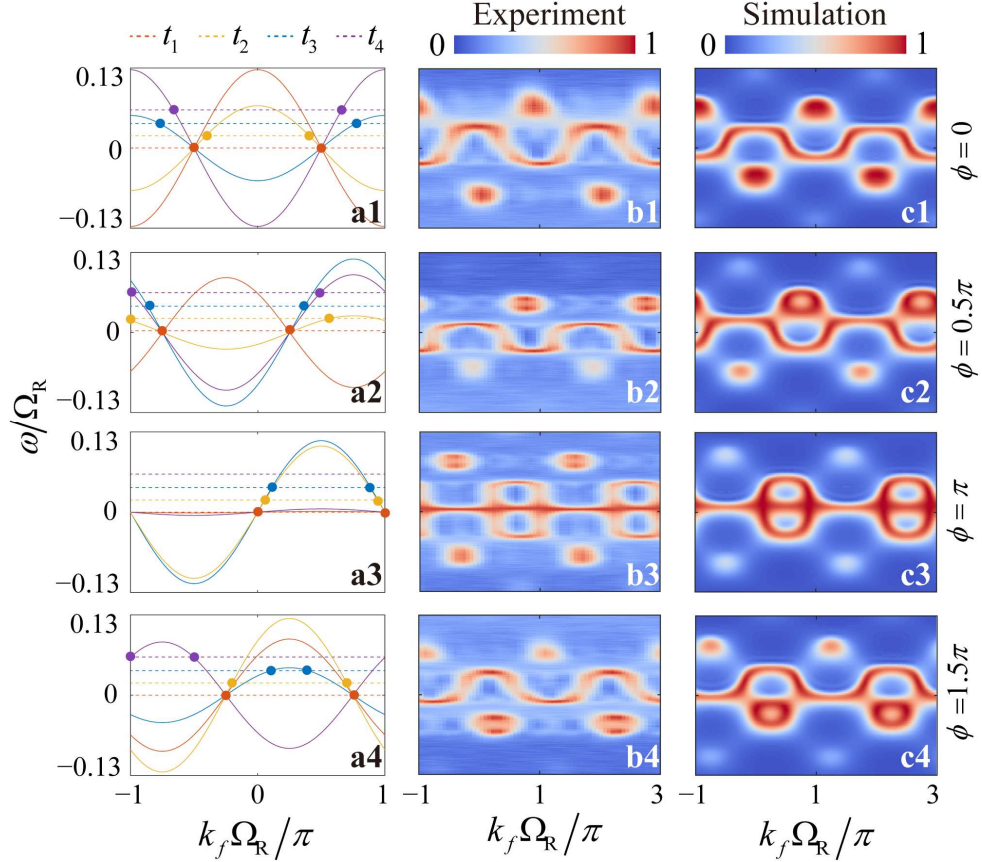

**Fig. 4. a1-a4,** The evolution of the static band structures of the system in different phase differences $\phi$ with time varying from $t_1$ to $t_4$. The colored circles label the intersection points of the excitation frequency (dashed lines) and the corresponding static band structure at different times. **b1-b4,** Measured trajectories of the dynamic band structures under modulation $V(t) = V_0\cos[(\Omega_R - \Delta_{\text{exp}})t] + V_0\cos[(\Omega_R + \Delta_{\text{exp}})t + \phi]$, with varied phase difference $\phi$ and $V_0 = 2.5$ V, $\Delta_{\text{exp}} = 5$ kHz. **c1-c4,** Simulated dynamic band structures under modulation $J(t) = g\cos[(\Omega_R - \Delta)t] + g\cos[(\Omega_R + \Delta)t + \phi]$, with different frequency difference $\phi$, and $g = 0.4$, $\Delta = 0.1$, $\eta = 0.002$.

Our chip can provide further opportunity for studying more complicated models with various time-dependent Hamiltonians. To show such cases, we rewrite the band structure in Eq. (4) to a general form

$$\varepsilon_{k_f}(t) = g_1\cos(k_f\Omega_R - \Delta_1 t) + g_2\cos(k_f\Omega_R + \Delta_2 t + \phi), \qquad (5)$$

which represents the superposition of two travelling-wave-feature band components. By separately manipulating either one of the two components in Eq. (5), one can obtain arbitrary designed band shapes, corresponding to versatile time-dependent Hamiltonians. In experiments, we first consider a simple case for the single near-resonant modulation with the RF signal having the form $V_1(t) = V_0\cos[(\Omega_R - \Delta_{\text{exp}})t]$, i.e., the second component in Eq. (5) is zero.

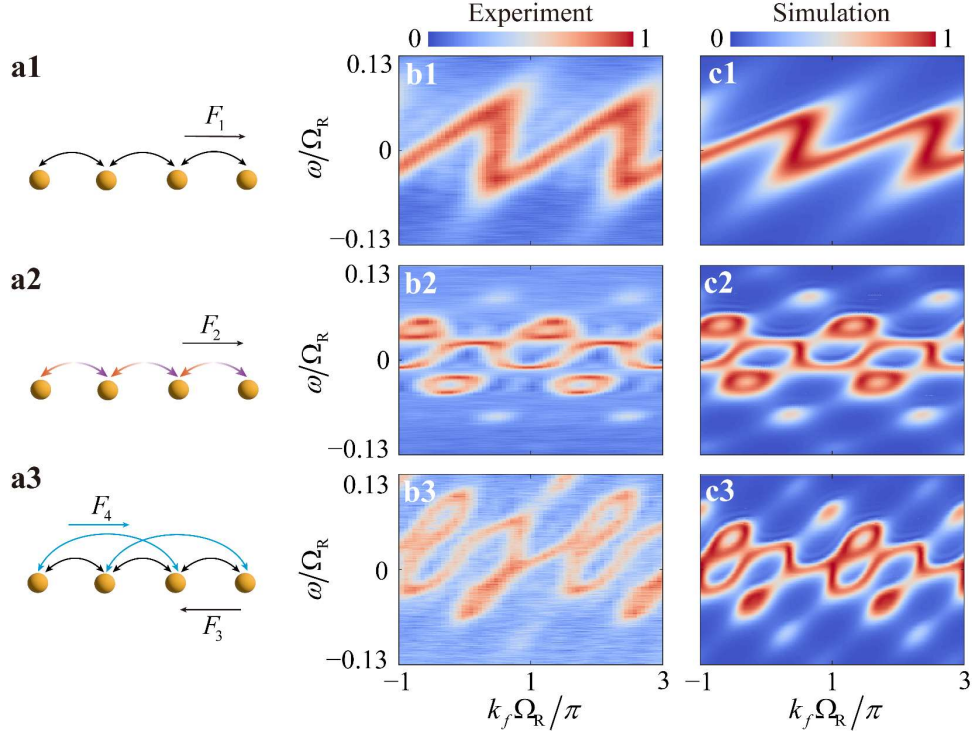

**Fig. 5.** Synthetic frequency lattice models and measured dynamic band structures under different combinations of RF signals: **a1-b1**, $V_1(t) = V_0\cos[(\Omega_R - \Delta_{exp})t]$, **a2-b2**, $V_2(t) = V_0\cos[(\Omega_R + \Delta_{exp})t] + V_0\cos[(\Omega_R - 2\Delta_{exp})t]$, and **a3-b3**, $V_3(t) = V_0\cos[(\Omega_R + \Delta_{exp})t] + V_0\cos[(2\Omega_R - 2\Delta_{exp})t]$, with $\Delta_{exp}$=5 kHz and $V_0 = 2.5$ V. The arrows in **a1-a3** denote the directions of the generated effective external forces. The simulated dynamic band structures under conditions **c1**, $J_1(t) = g\cos[(\Omega_R - \Delta)t]$, **c2**, $J_2(t) = g\cos[(\Omega_R + \Delta)t] + g\cos[(\Omega_R - 2\Delta)t]$, **c3**, $J_3(t) = g\cos[(\Omega_R + \Delta)t] + g\cos[(2\Omega_R - 2\Delta)t]$, with $g = 0.4$, $\Delta = 0.1$, and $\eta = 0.002$.

It leads to the band structure $\omega_1(t) = V_0\cos(k_f\Omega_R - \Delta_{exp}t)$, corresponding to the model of a one-dimensional photonic lattice under an external force $F_1$ [see Fig. 5a1][35]. One sees that the dynamic band structure shifts over time along the $k_f$ axis, as illustrated in the experimental measurement in Fig. 5b1 and the numerical simulation in Fig. 5c1. As a step forward, we then choose the bichromatic RF signal in the form of $V_2(t) = V_0\cos[(\Omega_R + \Delta_{exp})t] + V_0\cos[(\Omega_R - 2\Delta_{exp})t]$, composed of two near-resonant modulations with asymmetric and unbalanced frequency detunings. Such a configuration in the experiment can support a band structure $\omega_2(t) = 2V_0\cos(3\Delta_{exp}t/2) \cdot \cos(k_f\Omega_R - \Delta_{exp}t/2)$, which constructs a time-dependent lattice model with both the time-varying coupling strength and an effective external force $F_2$ [see Fig. 5a2]. It leads to the band structure oscillating along the vertical energy axis while shifting along the $k_f$ direction [see Figs. 5b2-c2 for experimental and numerical results]. We further tune the modulation form to $V_3(t) = V_0\cos[(\Omega_R + \Delta_{exp})t] + V_0\cos[(2\Omega_R - 2\Delta_{ex})t]$ combining both the nearest-neighbor coupling and next-nearest-neighbor (NNN) coupling near twice the number of FSR, which gives $\omega_3(t) = V_0\cos(k_f\Omega_R + \Delta_{exp}t) + V_0\cos(2k_f\Omega_R - 2\Delta_{exp}t)$. This choice of the modulation gives a one-dimensional lattice under two effective external forces $F_3$ and $F_4$ with opposite directions along NN and NNN couplings respectively [see Fig. 5a3], where the corresponding dynamic band structures from the experiment and simulation are plotted in Figs. 5b3-c3.

In conclusion, we demonstrate the photonic analog of various time-dependent Hamiltonian models in a single microresonator fabricated on a TFLN chip. By applying the bichromatic EO modulation with two near-resonant modulation frequencies oppositely detuned from the resonant frequency, a high-quality synthetic frequency lattice with time-varying coupling strength is constructed. Such lattice holds cosinusoidal-shape band feature with the band's amplitude periodically oscillating with time. Such time-dependent evolution is tracked by measuring the dynamic band structures in the experiment. By engineering the parameters of the bichromatic drive, versatile periodically modulated lattice models are achieved, which proves the capability of our chip for the on-demand simulations of time-dependent Hamiltonians. Our experimental work thus offers great possibilities in realizing more complex time-dependent Hamiltonian models associated with non-Hermitian physics and higher dimensionality in the future[46–49].

**Methods**

**Device fabrication**

The device was fabricated on 600-nm-thick x-cut TFLN on a 2-μm silicon dioxide buffering layer and a 500-μm silicon substrate (NanoLN). The bus waveguide and racetrack microresonator was patterned with standard electron-beam lithography. The patterns are then transferred into the TFLN layer in an inductively coupled plasma (ICP) step using argon (Ar+) plasma etching. The etch depth is 350 nm. The device is then cleaned by buffered HF solution and RCA1 cleaning solution. Gold electrode patterns are then created using laser direct writing, and the metal (15 nm of chromium and 300 nm of gold) is deposited using thermal evaporation and the bilayer lift-off process. The chip is then diced and the facets are polished for end-fire optical coupling. The air-cladded TFLN racetrack microresonator has a top width of 1.4 μm and a sidewall angle of 60 degree. It is near the critical coupling condition with the bus waveguide. The fiber-to-chip coupling loss in the telecom band is approximately 6 dB per facet.

**Experimental Setup**

A continuous wave laser source with pump power of 25 mW is utilized, whose polarization is adjusted by a polarization controller [see Fig. 2a]. The light filed is coupled through a lens fiber aligned with the end face of the bus waveguide. Two RF signals generated by two microwave sources with 40 GHz bandwidth are coupled by a microwave coupler and then applied on the patterned electrodes through the electronic probe. Two RF amplifiers with amplification gain 20 dB are used to boost the two RF signals. A phase shifter with 18 GHz bandwidth is placed after one RF amplifier to adjust the phase difference ($\phi$) between the two RF signals. The light signal coupled out from the through port of the bus waveguide by a second lens fiber is send to an optical spectrum analyzer for measuring the output spectra in Figs. 2b and 2d. As for the band structure measurements in Figs. 3-5, the light signal is sent to a fast photodiode with 40 GHz bandwidth for detection, and then recorded by an oscilloscope (16 GHz bandwidth and 80 GS/s sampling rate). To measure the second-order group velocity dispersion parameter of the microresonator ($D_2/2\pi$) in Fig. 2c, a 1×2 fiber coupler couples 5% of the input light source to a fiber Mach–Zehnder interferometer (MZI), and then sent to the photodiode. By simultaneously collecting the transmissions from the MZI and the microresonator while scanning the pump laser, and calibrating the wavelength by unwrapping the phase in the MZI transmission spectrum, one can measure the cavity-mode frequency with precision on the order of MHz.

**Acknowledgements**

The research is supported by National Natural Science Foundation of China (12104297, 12122407, 12192252, 62022058, 12293052, and 11934012), and National Key Research and Development Program of China (2023YFA1407200, 2022YFA1205101, and 2021YFA1400900). Y.Z. and L.Y. also thank the sponsorship from Yangyang Development Fund. This work was partially carried out at the USTC Center for Micro and Nanoscale Research and Fabrication.